\documentclass{aa}
\usepackage{graphicx}
\usepackage{txfonts}
\usepackage{natbib}
\bibpunct{(}{)}{;}{a}{}{,} 

\begin{document}

\title{A 100\,ks XMM-Newton view of the Seyfert 1.8 ESO\,113-G010. } 
\subtitle{I. Discovery of large X-ray variability and study of 
 the Fe\,K${\alpha}$ line complex.}
\author{D.\ Porquet\inst{1} 
\and P. Uttley\inst{2}
\and J.~N.\ Reeves\inst{3,4,5}
\and A.\ Markowitz\inst{3,4}
\and S.\ Bianchi\inst{6}
\and N.\ Grosso\inst{1} 
\and L.~Miller\inst{7}
\and S.\ Deluit\inst{8}
\and I.~M.~\,George\inst{9}
}

\offprints{D. Porquet}

\institute{
Observatoire Astronomique de Strabsourg, Universit\'e Louis Pasteur,
CNRS, 11 rue de l'Universit\'e, 67000 Strasbourg, France
\newline
\email{porquet@astro.u-strasbg.fr}
\and School of Physics and Astronomy, University of Southampton,
Southampton SO17 1BJ, UK 
\and Laboratory for High Energy Astrophysics, Code 662, 
NASA Goddard Space Flight Center, Greenbelt, MD 20771, USA
\and Dept. of Physics and Astronomy, Johns Hopkins University, 3400 North
Charles Street, Baltimore, MD 21218, USA
\and Astronomy Group, School of Geographical and Physical Sciences, 
University of Keele, Keele, Staffordshire, ST5 5BG, UK. 
\and Dipartimento di Fisica, Universit\`a 
degli Studi Roma Tre, via della Vasca Navale 84, 00146 Roma, Italy  
\and Dept. of Physics, University of Oxford, Denys Wilkinson Building,
Keble Road, Oxford OX1 3RH, UK
\and Centre d'Etude-Spatiale des Rayonnements, 9 Avenue du Colonel
 Roche, BP 4346, 31028 Toulouse Cedex 4, France
\and Dept. of Physics, University of Maryland Baltimore County, 
1000 Hilltop Circle, Baltimore, MD 21250, USA
}

\date{Received  / Accepted }

\abstract
{The Seyfert 1.8 galaxy ESO\,113-G010 had been observed 
  for the first time above 2\,keV by XMM-Newton during a short
  exposure ($\sim$4\,ks) in May 2001. 
In addition to a significant soft X-ray excess, it showed one of the
strongest (in EW) redshifted Fe\,K$\alpha$ lines, at 5.4\,keV.}
{We present here a long (100\,ks) XMM-Newton 
follow-up of this source performed in November 2005, in order to study
over a longer time-scale its main X-ray properties.}
{We use both timing analysis (Power Spectra Density analysis, rms
  spectra, flux-flux analysis) 
  and spectral analysis which mainly focuses on the Fe\,K${\alpha}$ line
  complex.} 
{The source was found in a higher/softer time-averaged flux state, 
and timing analysis of this source reveals strong, rapid
variability. The Power Spectral Density (PSD)
analysis indicates (at 95$\%$ confidence level)
a break at 3.7$^{+1.0}_{-1.7} \times 10^{-4}$~Hz.
This cut-off frequency is comparable to those measured
in some other rapidly-variable Seyferts, such as MCG--6-30-15
and NGC\,4051. From the mass-luminosity-time-scale, we infer
 that $M_{\rm BH}$ ranges from
$4\times10^{6}$--$10^{7}$~M$_{\odot}$ and the source is accreting at or 
close to the Eddington rate (or even higher). 
The existing data cannot
distinguish between spectral pivoting of the continuum and a
two-component
origin for the spectral softening, primarily because the data do not span
a broad enough flux range. In the case of the two-component model,
the fractional offsets measured in the
flux-flux plots increase significantly toward higher energies
(similar to what is observed in MCG--6-30-15)
as expected if there exists a constant reflection component.
Contrary to May 2001, no significant highly redshifted emission line is
  observed (which might be related to the source flux level), while
  two narrow emission lines at about 6.5\,keV and 7\,keV are 
  observed. The S/N is not high enough to establish if the lines are
  variable or constant. 
As already suggested by the 2001 observation, no
significant constant narrow 6.4\,keV Fe\,K${\alpha}$
line (EW$\leq$32\,eV) is observed, hence excluding any dominant emission from
distant cold matter such as a torus in this Seyfert type 1.8 galaxy.}
{}


\maketitle

\section{Introduction}

The Seyfert ESO113-G010 (z=0.0257) was observed for the first time in 
X-rays, in December 1995, by {\sl ROSAT}, revealing an X-ray
luminosity of about 2.5 $\times$ 
10$^{43}$\,erg\,s$^{-1}$ (assuming $H_{\rm
  0}$=75\,km\,s$^{-1}$\,Mpc$^{-1}$  
and $q_{\rm 0}$=0.5) and some indication of X-ray variability
\citep{Pietsch98}. 
An optical follow-up (2.2 m ESO/MPG telescope at la Silla) performed 
in November 1996 done by the
same authors to characterize this bright galaxy led to the
conclusion that this Seyfert is of type 1.8, 
 and measured log\,$L_{FIR}$=44.34\,erg\,s$^{-1}$,
  log\,$L_{\rm B}$=43.97\,erg\,s$^{-1}$, and
  log\,$L_{\rm X,ROSAT}$=43.40\,erg\,s$^{-1}$. 
This object was observed for the first time above 2\,keV with {\sl
  XMM-Newton} on May 3, 2001 during a short exposure of about 4\,ks.
That observation revealed a soft excess below 0.7\,keV and, more interestingly,
a narrow emission line at 5.4\,keV (rest-frame), most probably
originating from a redshifted iron Fe\,K$\alpha$ line \citep{P04b}. 
The line was detected at 99\% confidence, calculated via 
Monte Carlo simulations which fully account for the range of energies 
where a narrow iron line is likely to occur.  
Such narrow spectral features in the 5--6\,keV energy range 
were discovered with {\sl XMM-Newton} and {\sl Chandra} in several
other AGN: 
e.g.  \object{NGC 3516} \citep{Tu02}, 
\object{NGC 7314} \citep{Yaqoob03}, \object{Mkn 766} \citep{JTurner04},  
\object{AX J0447-0627} \citep{DellaCeca05}. 
The redshifted Fe\,K$\alpha$ line discovered in ESO\,113-G010 had an EW
of $\sim$ 265 eV, making it one of the strongest (in EW) redshifted
iron lines observed to date in an AGN. 
The energy of the redshifted line could indicate emission from 
relativistic (0.17--0.23\,c) ejected matter moving away from the observer,  
as proposed for Mrk 766 by \cite{JTurner04}. 
Alternatively, if the
classification as type 1.8 object is correct, 
then emission from a narrow annulus 
 at the surface of the accretion disc 
is unlikely due to the very small inclination angle 
 (i.e. less than 10$^{\circ}$) required to explain the narrow, 
redshifted line in this intermediate Seyfert galaxy. 
However, emission from a small, localized 
hot-spot on the disc, occurring within a fraction of a complete disc orbit,  
could also explain the redshifted line (e.g., \citealt{Tu02,Dovciak04}). \\

Here, we present a 100\,ks {\sl XMM-Newton} follow-up observation  
of ESO\,113-G010.  
In section~\ref{sec:data}, we present the long {\sl
  XMM-Newton} analysis performed in November 2005, and the corresponding data
  reduction used for this work. Section~\ref{sec:timing} presents the
  timing analysis: light curves, PSD analysis, rms-spectrum, and flux-flux
  analysis. Section~\ref{sec:spectra} presents the spectral analysis, 
focusing on the study of the Fe\,K${\alpha}$ line complex between 6.4--7\,keV. 
Finally, we summarize and discuss the main results of the present work 
(section~\ref{sec:discussion}). 

\section[]{XMM-Newton observation and data reduction}\label{sec:data}

{\sl XMM-Newton} observed ESO\,113-G010
starting on November 10, 2005 (orbit 1085) 
with exposure times of $\sim$\,103\,ks and $\sim$\,102\,ks for the MOS,
and PN respectively. 
The EPIC-MOS cameras \citep{Turner01}  
 operated in the Large Window mode, 
while the EPIC-PN camera \citep{Str01} 
 operated in the standard Full Frame Window mode. 
Background flaring was negligible during most of the observation and
therefore the whole observation was used, except when mentioned in
Section~\ref{sec:timing}.  
The EPIC data were re-processed and cleaned using the {\sl XMM-Newton} 
{\sc SAS version 6.5.0} (Science Analysis Software) package. 
The net resulting livetimes were 102.1\,ks and 90.6\,ks
respectively for the MOS and PN cameras. 
Since pile-up was negligible, X-ray events corresponding to patterns
 0--12 and 0--4 events (single and double pixels) 
were selected for the MOS and PN, respectively. Only good X-ray events
(with FLAG=0) were included. 
The low-energy cutoff was set to 300 and 200\,eV for MOS and PN, respectively. 
The source data were extracted 
using a circular region centered on
 the source position of diameter of 50$^{\prime\prime}$\footnote{For
 the spectral analysis ($\S$4) the extraction region
 radius was taken to be the same for MOS1 and MOS2
 (i.e. 25\,$^{\prime\prime}$), in order to co-add both spectra.},  
25$^{\prime\prime}$, and 35$^{\prime\prime}$ 
for MOS1, MOS2 and PN respectively. The diameter was smaller for MOS2
compared to MOS1 due to a dead pixel column about
26$^{\prime\prime}$ from the source in the MOS2 data. 
ESO\,113-G010 was by far the brightest X-ray source in this 
30$^{\prime}$ EPIC field-of-view. 
Background data were taken from box regions 
on the same CCD as the source (excluding X-ray point sources). 
The {\sc xspec v11.3} software package was used for spectral 
analysis of the background-subtracted spectrum 
using the response matrices and ancillary files derived from the SAS tasks 
{\sc rmfgen} and {\sc arfgen}.


\section{Timing Analysis}\label{sec:timing}

\begin{figure}[!t]
\resizebox{\hsize}{!}{\rotatebox{0}{\includegraphics{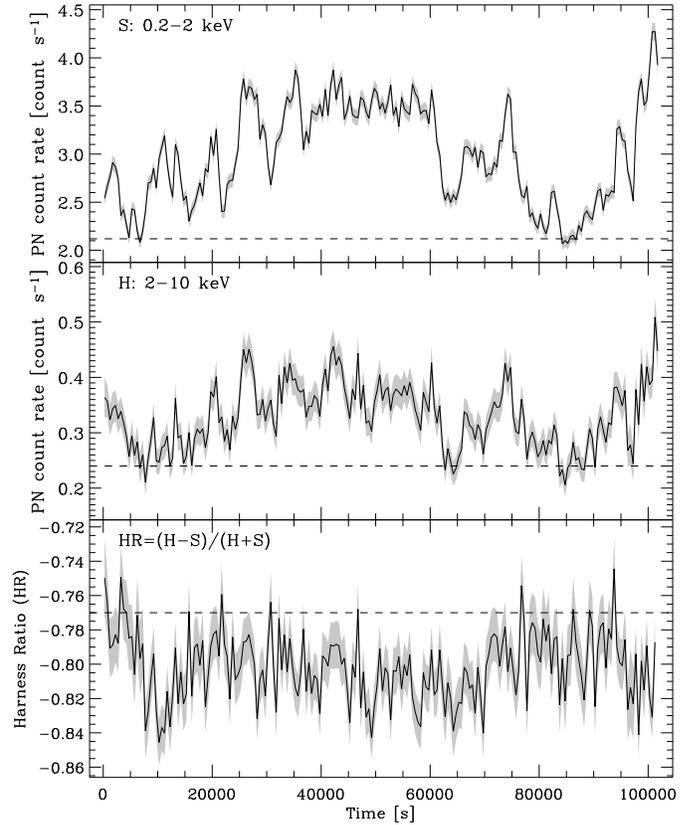}}}
\caption{Soft (S:\ 0.2--2\,keV) and hard (H:\ 2--10\,keV) 500-s
  resolution (background-subtracted) PN light curves of ESO\,113-G010
  (upper and middle panels, respectively), together with the hardness
  ratio between the two bands (lower panel). The grey stripes
  correspond to the 1$\sigma$ error bars. For comparison the count
  rate and hardness ratio level 
  found during the 4\,ks May 2001 observation are reported (dashed
  lines). }
\label{fig:HR}
\end{figure}

 PN light curves (time bin of 500\,s, and error bars of 1$\sigma$) clearly
  show fairly rapid variability with relatively large
  amplitude, both in the soft 
(S: 0.2--2\,keV) and hard (H: 2--10\,keV) energy ranges
 (Fig.~\ref{fig:HR}; upper and middle panels, respectively). 
The corresponding PN time-averaged count rate
  observed during May 2001 (duration of $\sim$\,4\,ks) 
is reported (Fig.~\ref{fig:HR},
dashed lines) and shows that the source was in a higher flux state for
virtually all of the 2005 observation when compared to the 2001
observation.  
The PN average count rates in the 0.2--12\,keV energy range 
are 2.42$\pm$0.02\,cts\,s$^{-1}$ and 3.71$\pm$0.01, for the 2001 and
  2005 observations, respectively.
The corresponding Hardness Ratio, HR ($\equiv$(H-S)/(H+S)) 
is also displayed (Fig.~\ref{fig:HR}, lower panel). 
The error bars for the hardness ratio are obtained using Gaussian
error propagation (see e.g., \citealt{Pietsch05}). The 2005 HR
  values show that the source was on average in a softer state 
 compared to May 2001. There
  was no very strong variation amplitude of the HR, except within the 
  first 10\,ks (with $<$$\Delta$HR$>\sim$0.07). \\

\begin{figure}[!Ht]
\resizebox{\hsize}{!}{\rotatebox{-90}{\includegraphics{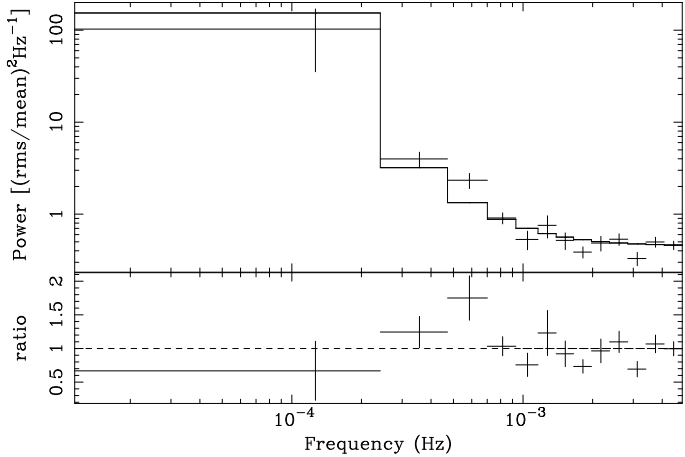}}}
\caption{0.3--10\,keV PSD of ESO~113-G010, including the ratio of the data to
the best-fitting power-law plus constant noise component model
(model shown as solid line).}
\label{psdpow}
\end{figure}

To quantify this
variability better, we carried out timing analysis using the power spectral
density function (PSD), the rms-spectrum, and flux-flux analysis.
To maximize the signal-to-noise we combined events from all three
instruments, re-scaling the background regions used for the MOS detectors
to match the source/background area ratio of the PN.  We used only data
from times between 8~ks and 95~ks from the start of the observation,
to remove the times of non-negligible background flaring for secure
timing analysis, and merged the good time
intervals of the different instruments in order to use
only data received simultaneously from all three instruments.  
The combined events file was used to extract light curves in different bands
to make the PSDs and rms-spectra.
Although the instrument responses differ, in all cases 
(including rms-spectra) we normalize by the mean flux in each band, so the
effect of the spectral response is taken out.

\subsection{PSD Analysis}\label{sec:PSD}

\begin{figure}[!Ht]
\resizebox{\hsize}{!}{\rotatebox{-90}{\includegraphics{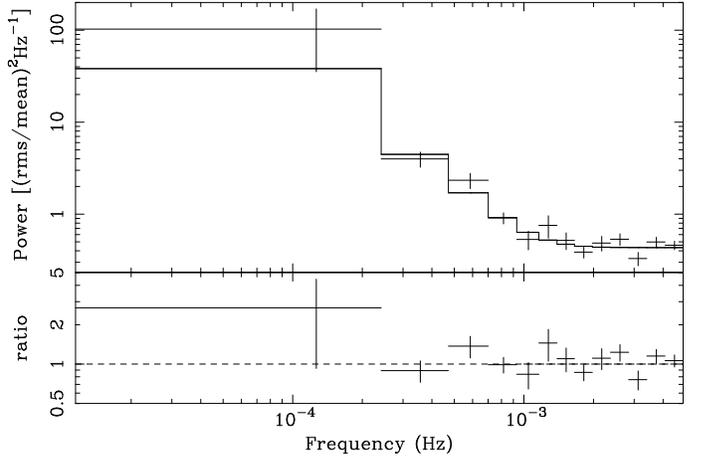}}}
\caption{0.3--10\,keV PSD of ESO~113-G010, including the ratio of the
  data to an 
exponentially cut-off power-law (with low-frequency slope
$\alpha=-1$), plus constant Poisson noise component (model shown as solid line).
}
\label{psdcutoff}
\end{figure}

We first consider the PSD for the light curve in the 0.3--10\,keV
energy range, which is shown in Fig.~\ref{psdpow}. 
 The PSD represents the average of the (squared) amplitude of
variations as a function of temporal frequency. 
The data are then fitted with a simple power law 
plus the flat level of variability due to
Poisson noise expected given the error bars.
A simple power-law plus noise is not a very good fit 
($\chi^{2}=24.1$ for 11 d.o.f.), and is rejected at $>98\%$
confidence (for power-law slope $\alpha=-2.1$).  
Allowing the noise level to be free improves the fit to
$\chi^{2}=18.9$ for 10 d.o.f., which is still rejected at $>95\%$ confidence.
Finally, leaving the noise level free while
allowing a break in the PSD improves the fit:
fixing the low-frequency slope to $-$1 (similar
to the low-frequency PSD shapes observed in other AGN, e.g. 
 \citealt{McHardy05}), we find a break frequency of
$(5.9\pm1.7)\times10^{-4}$~Hz (errors correspond to 90\% confidence 
for a single interesting parameter), with an upper limit on 
high-frequency slope of $\alpha<-2.9$.  The corresponding
$\chi^{2}=11.9$ for 9 d.o.f., which, according to the F-test, is a significant
improvement over the unbroken power-law model at the $95\%$ level.  
We also find a good fit ($\chi^{2}=13.9$ for 10 d.o.f.) for an exponentially
cut-off power-law, with low-frequency slope fixed to $-$1 and
cut-off frequency $(3.7^{+1.0}_{-1.7})\times 10^{-4}$~Hz, which we show
in Fig.~\ref{psdcutoff}.  

 We caution that the possible break or cut-off
in the PSD is only marginally significant, but note that, in support
of these models, the product of frequency and power, which is
constant on the slope=$-$1 part
of the PSD is $\sim$0.01, which is of the same order as that observed
in other AGN.  In comparison with other AGN, the measured cut-off
frequency is comparable to those measured in some other 
rapidly-variable Seyferts, such as MCG--6-30-15 ($8\times10^{-5}$~Hz,
 \citealt{McHardy05,Vaughan03a}), 
and  NGC~4051 ($8\times10^{-4}$~Hz, \citealt{McHardy04}).
 The correlation between AGN X-ray variability PSD break time-scale and 
black hole mass is now well-established. It appears that AGN PSD break 
time-scales can be linearly scaled down by the black hole mass
to match the breaks observed in the PSDs of stellar mass black holes in 
X-ray binary systems
(e.g. \citealt{Markowitz03,Uttley05}).  
Recently, McHardy et al. (2006) showed that the residual scatter in
the mass-time-scale relation could be explained if time-scales also scale with
    luminosity or equivalently, accretion rate. Combining AGN and stellar mass
    black hole data for Cyg~X-1, McHardy et al. framed the
    best-fitting relation 
    for break time-scale log$T_{\rm B} = 2.1 {\rm log} M_{\rm BH} -
    0.98 {\rm log} L_{\rm bol} - 2.33 $, with $T_{\rm B}$ 
    in units of days and $M_{\rm BH}$ and bolometric
    luminosity $L_{\rm bol}$ in units of $10^{6}$~M$_{\odot}$ and
    $10^{44}$~erg~s$^{-1}$ respectively. The best-fitting PSD break
    time-scale we 
    measure, 
$3.7\times10^{-4}$~Hz, corresponds to $T_{\rm B}=0.03$~days.  
For a range in $L_{\rm bol}$ from $10^{44.5}-10^{45}$~erg~s$^{-1}$ (i.e. 
assuming a minimum equal to the sum of FIR, optical and X-ray  
luminosities reported by \citealt{Pietsch98}, and in this work), we infer from the
mass-luminosity-time-scale 
relation of \cite{McHardy06} that $M_{\rm BH}$ ranges from
$4\times10^{6}$--$10^{7}$~M$_{\odot}$ and the source is accreting at or 
close to the
Eddington rate (assuming even higher $L_{\rm bol}$ implies that the AGN 
is a super-Eddington accretor).

We also measured PSDs using smaller subsets of the observed energy range,
but due to the relatively weaker $S/N$ of these PSDs, the energy
dependence of the PSDs could not be well constrained.

\subsection{The rms-spectrum}\label{sec:rms}
\begin{figure}[!Ht]
\resizebox{\hsize}{!}{\rotatebox{-90}{\includegraphics{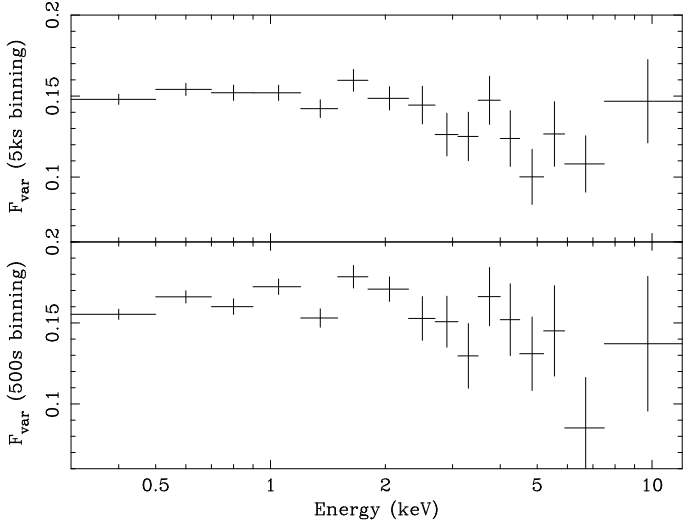}}}
\caption{The rms-spectra of ESO~113-G010, for light curve bin times of
  5~ks (top) and 500~s (bottom).  See text for details.
}
\label{rmsspec}
\end{figure}

The rms-spectrum quantifies the spectral shape of the variable
part of the emission (see \citealt{Vaughan03b} for
a review).  Specifically, one measures the
root-mean-square (rms) variability in light curves made from
a number of narrow energy bands or `channels', and 
then normalizes each rms value by the mean in that channel to
obtain the fractional rms (also known as $F_{\rm var}$).
A plot of $F_{\rm var}$ versus energy yields the shape of
the variable part of the spectrum relative to the mean spectrum
(i.e. independent of instrumental spectral response).  
The error bars are determined using the formula of 
 \cite{Vaughan03b}.  We show
the measured rms-spectra in Fig.~\ref{rmsspec}, 
obtained using light curves binned to 5~ks (top panel) and
500~s (bottom panel).  The rms-spectra obtained with 500~s
binning naturally includes faster fluctuations than for the 5~ks binned
data, and the $F_{\rm var}$ measurements are accordingly larger. 
$F_{\rm var}$ appears to decrease above about 2~keV for both
bin timescales. Therefore the soft X-ray excess down to 0.3\,keV 
appears more variable compared to the hard energy range. This is
different from the typical rms-spectra 
consistent with a broad peak in the variability amplitude between 0.8
and 2\,keV, and a much lower variability amplitude below 0.8 keV  
(e.g., \citealt{Fabian02,Inoue03,Ponti06}). 
In addition, there is a hint of a dip in the 500\,s bin
rms-spectrum around
the Fe\,K$\alpha$ complex energy ($\sim$6.4--7\,keV).
This could be due to a reflected component remaining constant
while the power-law continuum varies, as proposed for the two-component
model ($\S$3.3). An origin in a distant material is unlikely since no narrow
6.4\,keV Fe\,K${\alpha}$ line is observed (cf.\ $\S$~\ref{sec:spectra}).

\subsection{Flux-flux analysis}\label{sec:fluxflux}

Many AGN show a characteristic steepening of their
continuum X-ray spectra at higher fluxes.
A useful insight into the origin of this continuum spectral
variability can be obtained using flux-flux analysis.
 \cite{Taylor03} showed how a simple plot of soft X-ray
flux versus a harder X-ray flux can indicate whether spectral
variability is due to pivoting at high energies (where
the flux-flux plot takes a power-law form), or due to
changes in the strength of a continuum component with a constant
spectral shape plus a constant-flux, constant spectral shape
harder component
(where the flux-flux plot is linear with a positive intercept on
the hard flux axis).  Here we carry out a flux-flux analysis
of the ESO~113-G010 data, using the same combined MOS and PN
events data used for the rms-spectral analysis.

\begin{figure}[!Ht]
\resizebox{\hsize}{!}{\rotatebox{-90}{\includegraphics{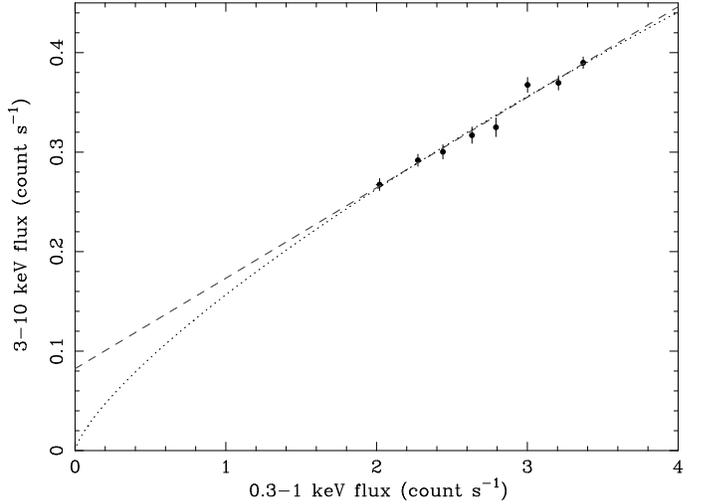}}}
\caption{3--10\,keV versus 0.3--1\,keV binned flux-flux plot of
  ESO~113-G010, compared 
with a best-fitting linear plus constant model (dashed line,
representing two-component spectral variation) and a simple power-law
(dotted line, representing high energy spectral pivoting).  The two models
are indistinguishable over the relatively narrow flux range covered by the
data.
}
\label{fluxflux}
\end{figure}

We first realize a flux-flux plot to compare the 0.3--1\,keV
and 3--10~keV ranges.  We measured the fluxes in simultaneous 500~s
bins, and following the approach of \cite{Taylor03},
we bin the 3--10~keV flux as a function of the 0.3--1~keV flux,
obtaining the standard error in the mean 3--10 keV~flux of each bin
from the spread in fluxes in the bin.  The resulting flux-flux plot is shown
in Fig.~\ref{fluxflux}.  A linear plus constant model provides a good
fit to the data, with $\chi^{2}=5.4$ for 6 d.o.f., for positive
constant offset on the 3--10~keV flux axis of
$0.082\pm0.014$~count~s$^{-1}$ (error bars are $1\sigma$).
A single power-law (with no offset)
also provides a good fit to the flux-flux plot
($\chi^{2}=5.4$ for 7 d.o.f.), for a power-law
index of $0.75\pm0.04$.  Thus, the existing data cannot
distinguish between spectral pivoting of the continuum and a two-component
origin for the spectral softening, primarily because the data span
an insufficiently broad range of fluxes 
to see the characteristic change in gradient
expected from the power-law.

\begin{figure}[!Ht]
\resizebox{\hsize}{!}{\rotatebox{-90}{\includegraphics{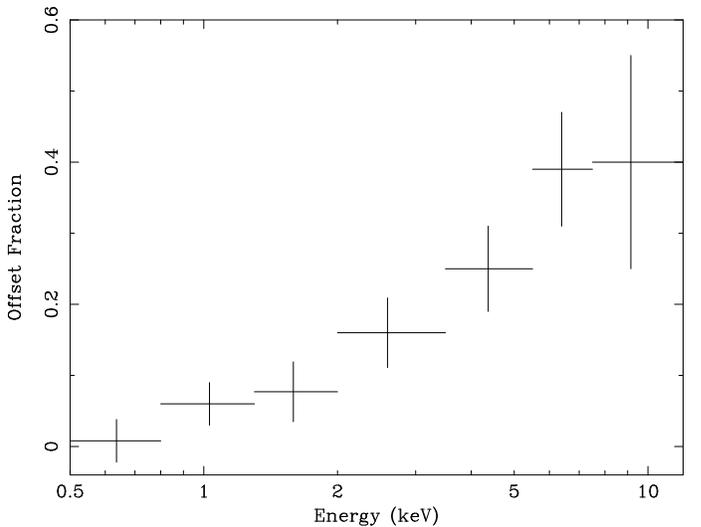}}}
\caption{Fractional offset spectrum, showing fractional constant offsets
(y-axis intercept divided by mean flux) for a linear plus constant model
fitted to various energy bands versus the 0.3--0.5~keV reference band.
In the case that the two-component model for spectral variability is correct,
the offset spectrum shows lower limits on the strength of
the constant component spectrum relative to the mean spectrum.}
\label{offsetspec}
\end{figure}

\begin{figure*}[!Ht]
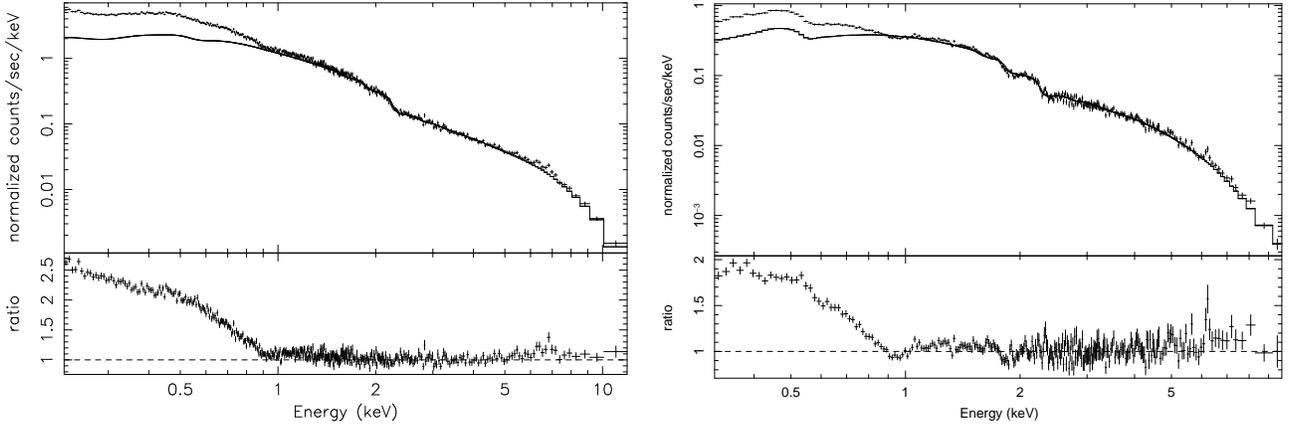

\begin{center}
\begin{tabular}{cc}
\resizebox{0.45\hsize}{!}{\rotatebox{-90}{\includegraphics{AA7699_f7l.ps}}}
&
\resizebox{0.45\hsize}{!}{\rotatebox{-90}{\includegraphics{AA7699_f7r.ps}}}
\end{tabular}
\end{center}
\caption{Spectra of ESO113-G010 (in the observer frame) observed in
  November 2005. A power-law has been fitted to the 2--5\,keV data
and extrapolated to lower and higher energies.
A significant soft X-ray positive residual is seen below 1\,keV, as well as
 the presence of a Fe\,K$\alpha$ complex near 6--7\,keV.
For presentation, the data have been re-binned into groups of 15 bins
for PN (10 for MOS),
after group of a minimum of 100 and 50 counts per bin were used for the
  fit for the PN and MOS respectively.
{\it Left panel:} 0.2--12\,keV PN spectrum. {\it Right panel:}
  co-added MOS 0.3--10\,keV spectrum.
 }
\label{fig:spectrum}
\end{figure*}

Although we cannot distinguish between spectral pivoting
and a two-component spectral model, we can use the flux-flux
method to determine the shape of any constant component under the
assumption that the two-component model is correct.  Following
the approach of \cite{Taylor03}, we fit linear plus constant
models to the flux-flux plots
measured from light curves obtained in a number of narrow bands.
For the $x$-axis, we use the 0.3--0.5~keV band as our `reference' band.
Thus all offsets measured in other bands are relative to this band
only, under the assumption that there is zero flux in the constant
component in the reference band.  If there is a non-negligible flux
in the reference band, the implied offsets in other bands increase
accordingly.  Thus, if the two-component model is correct
the offsets we measure from flux-flux plots here
represent a lower-limit on the strength of any constant component.
To remove the effects of
the instrument spectral response, we normalize each measured offset
by the mean flux in each band.  The resulting spectrum of fractional
offsets is shown in Fig.~\ref{offsetspec}.  Since
fractional offset increases significantly toward higher energies,
the offset spectrum
is harder than the mean spectrum, implying that any
constant component is also significantly harder than the variable component,
as expected if we are witnessing a constant reflection component.
Notably the reflected flux contributes about $40\%$ of the mean
at the hardest energies, similar to what is observed
in MCG--6-30-15 using the same approach \citep{Vaughan04}.

\section[]{Spectral analysis}\label{sec:spectra}

In this section, we present the spectral analysis of ESO\,113-G010. 
In this article, we focus on the analysis of the 2--12\,keV hard X-ray range
 (not affected by WA signatures), including the study of both the 
continuum and the Fe\,K${\alpha}$ line
complex between 6.4--7\,keV.  
We co-added the MOS1 and MOS2 data into a single spectral file to 
maximize the signal-to-noise ratio. 
The time-averaged PN and MOS spectra  are binned 
to give a minimum of 100 and 50 counts per bin, respectively; while the 
sub-spectra used in the time-resolved spectral analysis
(see below) are binned to give a minimum of 20 counts per bin. 
In all subsequent fits (above 2\,keV), we include the Galactic column density 
($N^{\rm Gal}_{\rm H}$=2.74 $\times 10^{20}$\,cm$^{-2}$), 
 obtained from the {\sc coldens} program using the compilations of
 \cite{DL90}. 
Note that all fit parameters are given in the source's rest frame, 
with values of $H_{\rm 0}$=75\,km\,s$^{-1}$\,Mpc$^{-1}$, 
and $q_{\rm 0}$=0.5 assumed throughout. 
The errors quoted correspond to 90$\%$ confidence ranges for one
 interesting parameter ($\Delta \chi^{2}$=2.71).
 Abundances are those of \cite*{An89}. 
In the following, we use the updated cross-sections for X-ray absorption by 
the interstellar medium ({\sc tbabs} in {\sc XSPEC}) from \cite*{Wilms00}.  

\subsection{The main spectral components}

As a first step, we fit an absorbed 
power-law model over the 2--5\,keV energy range where the spectrum should
be relatively unaffected by the presence 
of a broad soft excess, a warm absorber-emitter 
medium, an Fe\,K${\alpha}$ emission line, 
and a contribution above 8\,keV from 
a high energy Compton reflection hump.
In this energy range, the PN data are well fitted by a single power-law model
 with  $\Gamma$= 2.02$\pm$0.04 ($\chi^{2}$/d.o.f.=228.3/214).  
For comparison the value from the 2001 observation is 1.96$\pm$0.22.
This power-law index is consistent with values found for broad-line
radio-quiet quasars: 
 $\Gamma$=1.90 with a standard deviation of 0.27 (\citealt{P04a}). 
 The 2--10\,keV time-averaged luminosity (inferred from the parameter fit
  reported in Table~\ref{tab:linemean}) is
  4.0$\times$10$^{42}$\,erg\,s$^{-1}$.      
Figure~\ref{fig:spectrum} displays the 2--5\,keV PN spectrum 
extrapolated over the 0.2--12\,keV broad band energy. 
A strong positive residual is seen below 1\,keV 
due to the presence of a soft X-ray excess;  
and in addition, there is a positive residual in the 6--7\,keV energy
range (see \S~\ref{sec:FeK}).
For comparison, we also report the co-added MOS\,1 and MOS\,2 spectrum, which
confirms the presence of a complex soft excess as well as a positive
deviation near the 6--7\,keV. We henceforth focus only on the PN data, 
since the PN CCD has a better sensitivity over a broader energy range
 (0.2--12\,keV) compared to the MOS CCD. However we have checked that 
all the spectral fits are consistent with those obtained with the MOS, 
albeit with much lower photon statistics. 
We would like to note that the RGS data show the presence of several
narrow blue-shifted absorption lines that reveals the presence of a
warm absorber (WA) in outflow. The presence of this medium has to be
taken into account  
in order to fit properly the overall 0.2--12\,keV PN spectrum.
Spectral analysis combining RGS and 
PN data will be presented in a forthcoming paper.

\begin{table*}[!Ht]
\caption{Best-fitting spectral parameters of PN time-averaged spectrum 
     in the 2--12\,keV energy range
 with an absorbed (Galactic, $N_{\rm H}$=2.74$\times$10$^{20}$\,cm$^{-2}$) 
power-law (PL) component plus a line profile: 
{\sc zgauss}: Gaussian profile; and {\sc diskline}: profile line
emitted by a relativistic accretion disc for a non-rotating black hole
 \citep{Fabian89} assuming an emissivity law $q$ equal to $-$2. 
(a): $R_{\rm out}$=1\,000\,$R_{\rm g}$.  (b): $R_{\rm
     out}$=1.2 $\times$ $R_{\rm in}$.  
The line fluxes are expressed in units of
     10$^{-6}$\,ph\,cm$^{-2}$\,s$^{-1}$. 
 F-test probabilities are calculated with the corresponding power-law
     model as reference. Fixed parameters are indicated by (f).
}
\begin{center}
\begin{tabular}{llllcllllllll}
\hline
\hline
\noalign {\smallskip}                       
{\small Model}      &  \multicolumn{1}{c}{\small $\Gamma$} &\multicolumn{6}{c}{\small Line parameters}&{\small $\chi^{2}$/d.o.f.} &   F-test   \\
\noalign {\smallskip}                       
                 &                    &  \multicolumn{1}{c}{E}    &
\multicolumn{1}{c}{$\sigma$} & \multicolumn{1}{c}{$R_{\rm in}$}       &
          \multicolumn{1}{c}{$\theta$}  &  \multicolumn{1}{c}{Flux}  & \multicolumn{1}{c}{EW} \\
\noalign {\smallskip}                       
                 &                    &  \multicolumn{1}{c}{
            (keV)}    &  \multicolumn{1}{c}{(eV)} &\multicolumn{1}{c}{($R_{\rm g}$)}  &
          \multicolumn{1}{c}{(deg)} & \multicolumn{1}{c}{} & \multicolumn{1}{c}{(eV)} \\
\noalign {\smallskip}                       
\hline
\hline
\noalign {\smallskip}  
  {\small PL }       & {\small 1.95$\pm$0.02}&   \multicolumn{1}{c}{ -- } &   \multicolumn{1}{c}{ -- } &   \multicolumn{1}{c}{ -- } &   \multicolumn{1}{c}{ -- } &   \multicolumn{1}{c}{ -- } &   \multicolumn{1}{c}{ -- } &     {\small 327.9/294} & \multicolumn{1}{c}{ -- }\\ 
\noalign {\smallskip}                       
{\small PL + 2$\times$zgauss}       & {\small 1.97$\pm$0.02} & {\small
  6.50$^{+0.06}_{-0.05}$} &  10 (f)  &     \multicolumn{1}{c}{ -- } &
\multicolumn{1}{c}{ -- } &   {\small 1.6$\pm$0.7}   &  {\small 52$\pm23$}     &         {\small 294.7/290}&  $>$ 99.99$\%$ \\ 
                        &       &                     {\small
  7.00$\pm$0.04} & 10 (f)  &    \multicolumn{1}{c}{ -- } &\multicolumn{1}{c}{ -- } & {\small 2.0$\pm$0.7}   &  {\small 73$\pm$26}     &       &  \\ 
\noalign {\smallskip}                       
{\small PL + {\sc diskline$^{(a)}$}} & {\small
  2.00$\pm$0.02} & {\small 6.75$\pm$0.08} &    \multicolumn{1}{c}{ --
  }   & {\small 6 (f)}
  & {\small $\geq$50} &{\small 7.7$^{+2.3}_{-2.6}$} & 305$^{+90}_{-104}$ &  {\small 299.3/291} & $>$99.99$\%$ \\
\noalign {\smallskip}                       
{\small PL + {\sc annulus$^{(b)}$}} & {\small
  1.97$\pm$0.02} & {\small 6.79$\pm$0.05} &    \multicolumn{1}{c}{ --
  } & {\small 191$^{+71}_{-63}$ }  & {\small 45 (f)} &{\small 7.7$^{+2.3}_{-2.6}$} & 159$^{+50}_{-49}$ &  {\small 299.1/291} & $>$99.99$\%$ \\
\noalign {\smallskip}                       
\hline
\hline
\end{tabular}
\end{center}
\label{tab:linemean}
\end{table*}

\begin{figure}[!Ht]
\resizebox{0.98\hsize}{!}{\rotatebox{-90}{\includegraphics{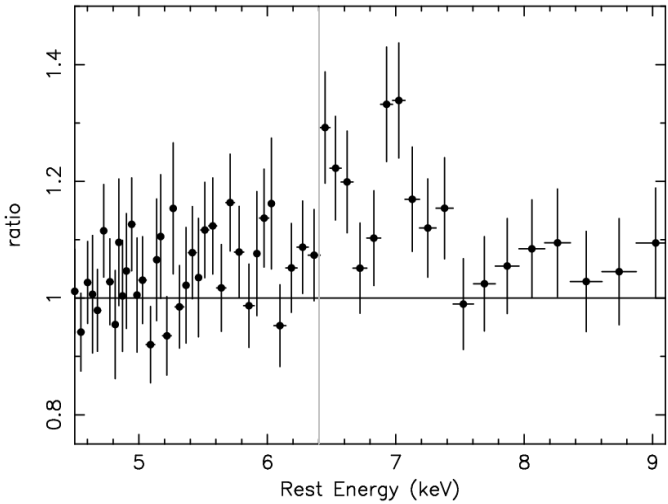}}}
\caption{Fe\,K line profile (quasar frame) 
when PN time-averaged data are fit with a power-law continuum. 
For presentation, the data have been re-binned into groups of 10 bins
for PN, after group of a minimum of 100 counts 
per bin were used for the fit. The grey vertical line corresponds to
6.4\,keV. 
}
\label{fig:meanFeK}
\end{figure}

\subsection{The Fe\,K${\alpha}$ line complex}\label{sec:FeK}

\subsubsection{Time-averaged analysis}\label{sec:average}
The positive residuals between 6.5--7\,keV in the quasar frame can 
be most probably associated with a Fe\,K$\alpha$ complex 
as observed in most AGN \citep{P04a,Pi05,Ji05}. The mean iron line
profile as a ratio of the continuum above 2\,keV is displayed in
Fig.~\ref{fig:meanFeK}, and shows a double-peaked line profile. 
Therefore, we fit the 2--12\,keV energy range with an absorbed
power-law continuum and two narrow Gaussian line profiles 
(Table~\ref{tab:linemean}). 
We found a much better fit,  with $\Delta\chi^{2}$=33.2 for 4 additional
parameters (F-test $>$ 99.9$\%$). 
The line energies are 6.50$^{+0.06}_{-0.05}$\,keV and 7.00$\pm$0.04\,keV. 
The F-test probabilities for each line are 99.7$\%$ and 99.98$\%$ for the
6.50\,keV and 7.0\,keV lines, respectively. 
Allowing the line widths to vary does not significantly
  increase the goodness of fit, with $\Delta\chi^2$ = 0.5
  for two additional parameters. 

The former energy is compatible within the error
 bars with a line emitted by moderately ionized iron
($\sim$\ion{Fe}{xix}--\ion{Fe}{xxii}). However this line is more
  likely a blue or red peak of a relativistic line, indeed a genuine
  \ion{Fe}{xix}--\ion{Fe}{xxii} (L-ions) emission line complex would suffer
  from resonant Auger destruction \citep{Ross96,Liedahl05}.  
The latter line energy corresponds to highly ionized iron line 
(\ion{Fe}{xxvi}) or to a highly blue-shifted iron line from lower ionization matter. 
However there is no evidence for narrow emission from neutral iron, 
the 90\% confidence upper-limit to a narrow ($\sigma$=10\,eV) iron
line at 6.4\,keV 
(a 6.5\,keV narrow line with a width of 10\,eV is also fitted to take
into account its 
emission contribution) is only 32\,eV. If the 6.5\,keV line
  contribution is not taken into account, we would infer
  $EW{\rm (6.4\,keV)}\leq$\,47eV. 
From the branching ratio of $\mathrm{Fe}~\mathrm{K}\beta /
  \mathrm{Fe}~\mathrm{K}\alpha$, expected to be less than 0.145 
(\citealt{Palmeri03}, and references therein), the maximum
  contribution of the Fe\,K$\beta$ (from cold iron) 
to the 7.00\,keV line is only up to 5\,eV.

We also tested if these two features can be explained by a double-peaked 
relativistic line profile emitted by a relativistic accretion disc 
for a non-rotating black hole ({\sc diskline}; \citealt{Fabian89}). 
The improvement of the fit is also significant (F-test $>$ 99.9$\%$)
and the line energy is about 6.8\,keV, i.e., corresponding to a highly
ionized iron ion 
(\ion{Fe}{xxv}--\ion{Fe}{xxvi}). 
Then, we checked whether the double-peaked like profile can be
  explained by emission from a disc annulus (using the {\sc diskline}
  profile with the width fixed to 20$\%$ of $R_{\rm in}$, i.e. $R_{\rm
    out}$=1.2 $\times$ $R_{\rm in}$). We fixed
  the disc inclination to 45$^{\circ}$ (representative of an
  intermediate type Seyfert). We found a very good fit to the data
  with a line energy of 6.79$\pm$0.05\,keV and an inner annulus radius
  of 191$^{+71}_{-93}$\,$R_{\rm g}$, indicating that the line is
  not emitted in the inner part of the disc. 
In the case where the
disc inclination is not fixed to 45 degrees, then the line solution is
degenerate, whereby the inner annulus radius increases with the disc
inclination, while the line energy decreases.
 
\begin{figure}[!Ht]
\resizebox{\hsize}{!}{\rotatebox{0}{\includegraphics{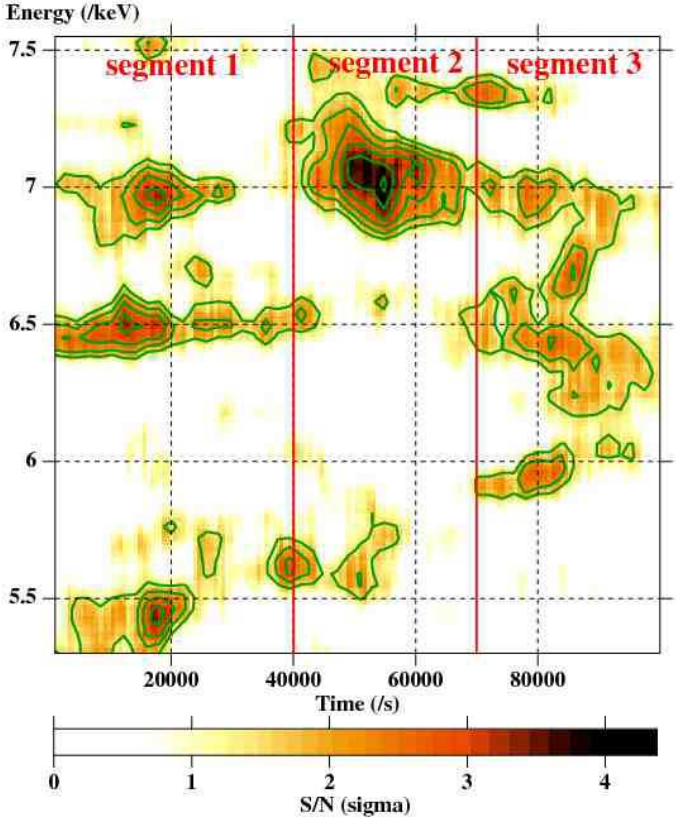}}}
\caption{Energy-time map. The energy values are in the quasar 
  frame. The contours 
represent 1.5--4.0 sigma above the continuum (in 0.5 increments), with 
darker bins representing an excess above the continuum.
 The segments 1-3 (delimited by red vertical lines) correspond to the
  time selections used for the time-resolved spectroscopic analysis.  
See text for explanation.}
\label{fig:SN}
\end{figure}

 Monte Carlo simulations were done to assess the
   significance of detecting the emission complex
   in this time-averaged spectrum. These simulations,
   described in detail in the Appendix, demonstrate that 
   regardless of the model used,
   the emission complex is indeed significantly detected.

We also checked for an upper limit to a line at 5.39\,keV,
  since such a line was found in the 2001 {\sl XMM-Newton} observation
  of May 2001. 
  We fixed the energy of the line to 5.39\,keV (see below), 
and the line width to 0.1\,keV (as done in
  \citealt{P04b}). We find EW$\leq$22\,eV (at 90$\%$ confidence level for
  one interesting parameter).\\

For comparison, we have re-processed the {\sl XMM-Newton} May
  2001 observation  with the same version of the SAS  (SAS 6.5.0) used
  for the 2005 observation. 
We fit the data with an absorbed (Galactic column density) power-law in the
  2--10\,keV energy range and four Gaussian lines in
  order to take into account the emission contribution of the lines
  near/at 5.4\,keV, 6.4\,keV, 6.5\,keV, and 7.0\,keV. 
The three latter line energies are fixed as well as their widths to
  10\,eV. Only the energy of the first Gaussian line is let free to
  vary  with the width fixed to 0.1\,keV (as done in \citealt{P04b}). 
One should notice that in \cite{P04b}, only the line at 5.4\,keV was
fitted. 
We found for the four lines: $EW_1$=246$\pm$147\,eV
($E_1$=5.39$\pm$0.09\,keV), $EW_2\leq$190\,eV ($E_2$=6.4\,keV), 
$EW_3\leq$100\,eV ($E_2$=6.5\,keV), and $EW_4\leq$208\,eV ($E_2$=7.0\,keV). 
 The EW for the line at 6.4\,keV, 6.5\,keV, and 7.0\,keV are compatible 
with the values found during the 2005 observation. For the line at
$\sim$5.39\,keV, the EW are not compatible and would mean that if the
line detected during the 2001 observation is real, it is
transient.

\subsubsection{Time-resolved analysis}

To test for possible rapid shifts in the energy and/or flux of the Fe
K$\alpha$ emission 
complex, we have created X-ray intensity maps in
the energy-time plane using the PN data. 
Photons from the source cell were accumulated in pixels in the
energy-time plane. The pixel distribution was smoothed in energy by the
instrumental resolution, using a Gaussian of 140\,eV
(appropriate for the single plus double events with the latest
calibration), and smoothed in time using a top-hat function of width
 20\,ks. Each 
time-slice was background-corrected by subtracting a time-dependent
background spectrum measured in an off-source region on the same 
detector chip as the source. The source continuum was modeled as an
absorbed power-law, of variable amplitude and slope but time-invariant
absorption column density. This continuum was subtracted, leaving positive and
 negative residuals that comprise noise plus any emission or
 absorption components on top of the continuum. 
More information about this method is given in \cite{Tu06}.
The ``signal-to-noise'' (S/N) map presented in Fig.~\ref{fig:SN}
 is the ratio of the fluctuation amplitude to the calculated noise. 

The  highest S/N observed is 4 and there are several points of 
significant excess emission above the continuum.  However, we note here 
that the energy-time map highlights fluctuations above the continuum 
which may be real spectral features, but does not assess the 
significance of variations in those features, assuming they are real.  
The apparent fluctuations in the energy-time map could be consistent 
with the complex emission features detected in the time-averaged 
spectrum being {\it constant}, e.g. if the average significance in one 
independent
10~ksec segment is 2~sigma above the continuum, then we might expect 
statistical variations from 0--4 sigma in a 100~ksec observation.  \\

\begin{figure}[!Ht]
\resizebox{8cm}{!}{\rotatebox{-90}{\includegraphics{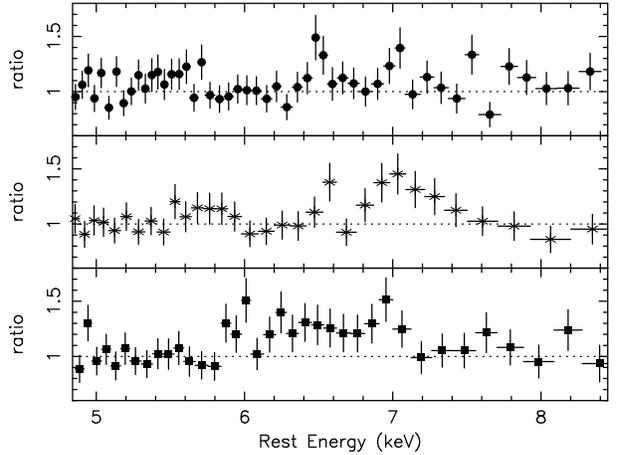}}}
\caption{The PN data/model sub-spectra of ESO\,113-G010 (quasar
  frame) for the three time intervals. 
The data have been fitted  assuming a
power-law underlying continuum. 
{\it Upper panel}: 0--40\,ks sub-spectrum (segment 1). 
{\it Middle panel}: 40--70\,ks sub-spectrum (segment 2). 
{\it Lower panel}: 70--100\,ks sub-spectrum (segment 3). }
\label{fig:segments}
\end{figure}

To quantify whether the line features are variable, we split the data 
set into three time intervals, 0--40~ksec, 40--70~ksec and 70--100~ksec, 
which are chosen to maximize the differences in the Fe~K line region 
which are highlighted by the energy-time map.  Note that in the case 
that the lines are constant, this approach essentially maximizes 
the effect of noise fluctuations that are picked up by the energy-time 
map, so that the statistical tests we carry out are in some sense the 
most optimistic for detecting line variability.  The line profiles for 
each segment (data/model ratio for a power-law continuum over the 
2--10\,keV energy range) are shown in Fig.~\ref{fig:segments}.
We then fit these three sub-spectra with (individual or double) 
Gaussian line profiles; the 
results are reported in Table~\ref{tab:line}. Though the line profiles
in Fig.~\ref{fig:segments} appear to be different, adding a double
Gaussian line profiles at 6.5\,keV and 7.0\,keV (as found in the
time-average spectrum) are significant (F-test
significance=99.5--99.97$\%$) in all segments. 
We also carried out Monte Carlo simulations to verify the significance 
of lines fitted to the individual sub-spectra; we detail these results 
in the Appendix (see \ref{app:trs}), but note here that they confirm
that the lines are  
significantly detected in the individual sub-spectra.

\begin{table*}[!Ht]
\caption{Best-fitting spectral parameters of PN time-resolved spectra 
     in the 2--10\,keV energy range
 with an absorbed (Galactic, $N_{\rm H}$=2.74$\times$10$^{20}$\,cm$^{-2}$) 
power-law (PL) component plus a Gaussian profile ({\sc zgauss}): . 
The line fluxes are expressed in 10$^{-6}$\,ph\,cm$^{-2}$\,s$^{-1}$. 
F-test probabilities are calculated with the corresponding power-law
     model as reference. (*) The F-test probability for freeing the line
     width in the fit is 92$\%$. (a) Monte-Carlo simulations have been
     performed (see text). Fixed parameters are indicated by (f).
}
\begin{center}
\begin{tabular}{lllllllc}
\hline
\hline
\noalign {\smallskip}                       
{\small Model}      &  \multicolumn{1}{c}{\small $\Gamma$} &\multicolumn{4}{c}{\small Line parameters}&{\small $\chi^{2}$/d.o.f.} &   F-test   \\
                 &                    &  \multicolumn{1}{c}{E}    &  \multicolumn{1}{c}{$\sigma$} &  \multicolumn{1}{c}{Flux}  & \multicolumn{1}{c}{EW} \\
\noalign {\smallskip}                       
                 &                    &  \multicolumn{1}{c}{
            (eV)}    &  \multicolumn{1}{c}{(eV)} & \multicolumn{1}{c}{} & \multicolumn{1}{c}{(eV)} \\
\noalign {\smallskip}                       
\hline
\hline
\noalign {\smallskip}                       
\multicolumn{8}{c}{0--40\,ks (segment 1)}\\
{\small PL }       & {\small 1.99$\pm$0.04}&   \multicolumn{1}{c}{ -- }   &    \multicolumn{1}{c}{ -- }       &   \multicolumn{1}{c}{ -- }   &    \multicolumn{1}{c}{ -- }       &     {\small 506.4/514} & \multicolumn{1}{c}{ -- }\\ 
{\small PL + zgauss$^{(a)}$}       & {\small 2.01$\pm$0.04} & {\small
  6.51$\pm$0.05} & {\small 10 (f)}   &  {\small 2.0$\pm$1.1}   &  {\small 66$\pm$36}     &   {\small 497.5/512}&   {\small 98.9$\%$} \\ 
{\small PL + zgauss$^{(a)}$}       & {\small 2.00$\pm$0.04} & {\small
  7.03$^{+0.21}_{-0.07}$} & {\small 10 (f)}   &  {\small 1.1$^{+1.1}_{-1.0}$}   &  {\small 45$^{+42}_{-38}$}     &   {\small 502.7/512}&    {\small 84.7$\%$} \\ 
{\small PL + 2$\times$zgauss}       & {\small 2.01$\pm$0.04} & {\small
  6.5 (f)} & {\small 10 (f)}   &  {\small 2.0$\pm$1.1}   &  {\small 68$\pm$37}     &   {\small 493.8/512}&   99.8$\%$ \\
{\small                   }       & {\small         } & {\small
  7.0 (f)} & {\small 10 (f)}   &  {\small 1.2$\pm$1.0}   &  {\small 48$\pm$41}     &   {\small  } &   \\ 
\noalign {\smallskip}                       
\hline
\noalign {\smallskip}                       
\multicolumn{8}{c}{40--70\,ks (segment 2)}\\
{\small PL }       & {\small 1.97$\pm$0.04}&   \multicolumn{1}{c}{ -- }   &    \multicolumn{1}{c}{ -- }       &   \multicolumn{1}{c}{ -- }  &   \multicolumn{1}{c}{ -- }   &     {\small 316.3/357} & \multicolumn{1}{c}{ -- }\\ 
{\small PL + zgauss}       & {\small 1.98$^{+0.05}_{-0.02}$} & {\small
  6.51 (f)} & {\small 10 (f)}   &  {\small $\leq$ 0.5}   &  {\small
  $\leq$ 54}     &         {\small 315.9/356}&   {\small 47.2$\%$} \\ 
{\small PL + zgauss}       & {\small 1.99$\pm$0.05} & {\small
 7.03$^{+0.07}_{-0.08}$ } & {\small 10 (f)}   &  {\small 3.2$\pm$1.4}   &  {\small
  125$\pm$54}     &         {\small 301.0/355}&   {\small 99.97$\%$} \\ 
{\small PL + zgauss$^{(a)}$}       & {\small 2.01$\pm$0.05} & {\small
  7.04$^{+0.10}_{-0.08}$} & {\small 142$^{+172}_{-104}$}   &  {\small 5.2$^{+2.9}_{-2.4}$}   &
{\small 204$^{+115}_{-92}$}     &         {\small 298.4/354}&
{\small 99.95$\%^{(*)}$} \\ 
{\small PL + 2$\times$zgauss}       & {\small 2.00$\pm$0.05} & {\small
  6.5 (f)} & {\small 10 (f)}   &  {\small $\leq$1.8}   &  {\small $\leq$58}     &   {\small 301.9/355}&   99.97$\%$ \\
{\small                   }       & {\small         } & {\small
  7.0 (f)} & {\small 10 (f)}   &  {\small 3.2$\pm$1.4}   &  {\small 124$\pm$54}     &   {\small  } &   \\ 
\noalign {\smallskip}                       
\hline
\noalign {\smallskip}                       
\multicolumn{8}{c}{70--100\,ks (segment 3)}\\
{\small PL }       & {\small 1.90$\pm$0.05}&   \multicolumn{1}{c}{ -- }   &    \multicolumn{1}{c}{ -- }       &   \multicolumn{1}{c}{ -- }   &    \multicolumn{1}{c}{ -- }       &     {\small 383.1/384} & \multicolumn{1}{c}{ -- }\\ 
{\small PL + zgauss}       & {\small 1.91$\pm$0.05} & {\small
  6.51 (f)} & {\small 10 (f)}   &  {\small 1.5$\pm$1.2}   &  {\small 51$\pm$42} &  {\small 379.1/383}&   {\small 95.5$\%$} \\ 
{\small PL + zgauss}       & {\small 1.91$\pm$0.05} & {\small
  7.03 (f)} & {\small 10 (f)}   &  {\small 1.5$\pm$1.2}   &  {\small 58$\pm$49} &  {\small 379.4/383}&   {\small 94.5$\%$} \\ 
{\small PL + zgauss$^{(a)}$}       & {\small 1.95$\pm$0.06} & {\small
  6.70$^{+0.27}_{-0.23}$} & {\small 431$^{+346}_{-183}$}   &  {\small 7.7$^{+5.1}_{-3.8}$}   &  {\small 273$^{+160}_{-134}$}     &         {\small 369.2/381}&   {\small 99.7$\%$} \\ 
{\small PL + 2$\times$zgauss}       & {\small 1.92$\pm$0.05} & {\small
  6.5 (f)} & {\small 10 (f)}   &  {\small 1.7$\pm$1.3}   &  {\small 57$\pm$43}     &   {\small 372.8/382}&  99.5$\%$ \\
{\small                   }       & {\small         } & {\small
  7.0 (f)} & {\small 10 (f)}   &  {\small 1.9$\pm$1.3}   &  {\small 76$\pm$51}     &   {\small  } &   \\ 
\noalign {\smallskip}                       
\hline
\hline
\end{tabular}
\end{center}
\label{tab:line}
\end{table*}

\begin{figure*}[!Ht]
\vspace{-0.5cm}
\begin{center}
\begin{tabular}{cc}
\resizebox{8cm}{!}{\rotatebox{-0}{\includegraphics{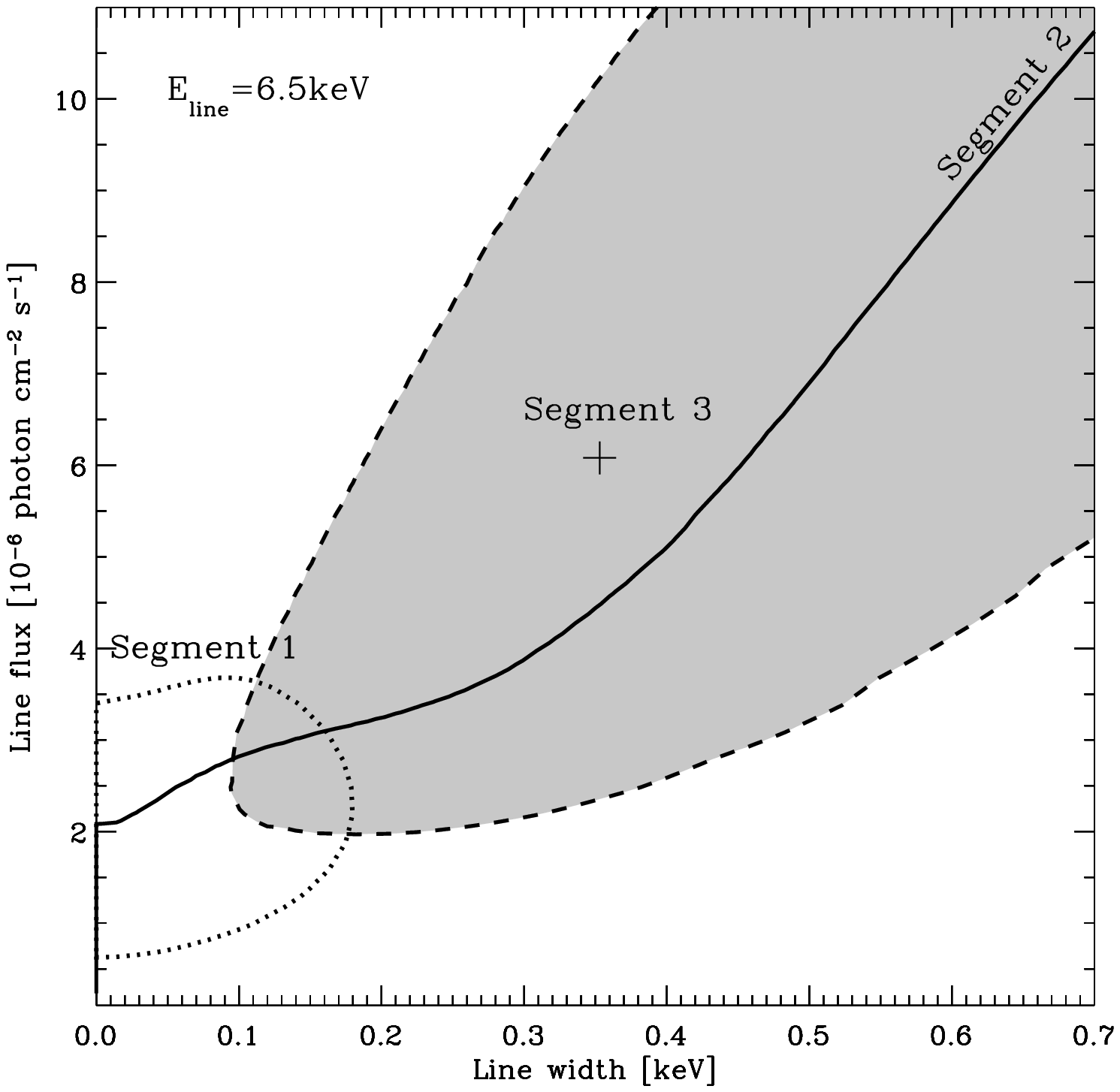}}} &
\resizebox{8cm}{!}{\rotatebox{-0}{\includegraphics{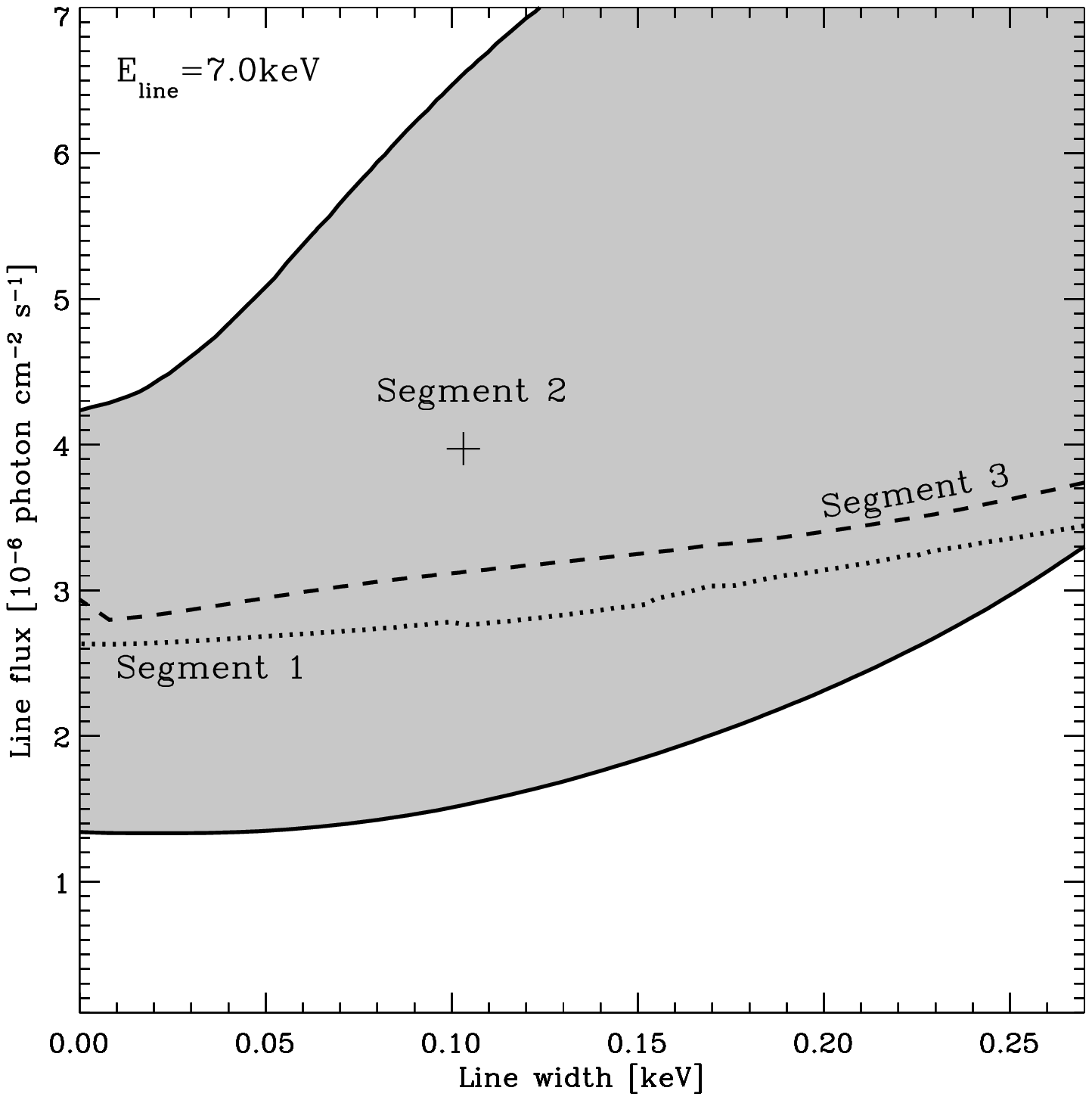}}}
\end{tabular}
\end{center}
\caption{Width and normalisation line contour plots at 90$\%$ confidence
  level ($\Delta\chi^{2}$=4.61) for segment 1 (dotted-line), segment 2
  (continuous line), and segment 3 (dashed-line). {\it Left panel}:
  line at 6.5\,keV. For segment 2, only an upper limit is found,
  corresponding to the continuous line. {\it Right panel}:
  line at 7.0\,keV. For segments 1 and 3, only an upper limit is found.}
\label{fig:contourgauss}
\end{figure*}

Having verified the statistical significance of the lines in each 
of the 3 segments, we check for significant variability of the
  line flux. 
 First, we fit simultaneously all segments with a power-law model plus two 
  Gaussian lines with line energies fixed to 6.5\,keV and 7.0\,keV and
  line widths 
  and line normalisations are free to vary within each segment. 
The contour plots at 90$\%$ confidence level for two
interesting parameters ($\Delta\chi^{2}$=4.61) are reported for each
line in Fig.~\ref{fig:contourgauss}, and show that assuming two Gaussian lines
 there are no statistically significant line variations.
 We also simulated 100 spectra using the {\sc fakeit} command,
    and assuming an initial model consisting of a power-law plus a single
    Gaussian emission line at either 6.5 or 7.0 keV. We assumed that
    the line is intrinsically constant, with parameters described by
    values, and used an exposure time of 29.8 ks. We found that
    the observed values of line intensity and line width $\sigma$
    were usually consistent with the resulting spread in 100
    simulated values. This suggested that we could not rule out
    at high significance the hypothesis that these parameters
    are intrinsically constant.

\section{Discussion}\label{sec:discussion}

We report the data analysis results of a long 100\,ks {\sl XMM-Newton}
observation in November 2005. 
We have used both model-independent techniques (e.g., rms-spectra,
flux-flux plots) and spectral analysis  
(time-averaged and time-resolved spectroscopy). 

\subsection{Timing analysis}

Light curves reveal that this source was 
highly variable on short timescales ($\sim$500\,s), and was on average
in a higher/softer state compared to the May 2001 observation.  
 The Power Spectral Density (PSD) 
analysis indicates (at 95$\%$ confidence)
a break at a frequency of 3.7($^{+1.0}_{-1.7})\times 10^{-4}$~Hz. 
This measured cut-off frequency is comparable to those measured 
in some other rapidly-variable Seyferts, such as MCG--6-30-15 
and NGC\,4051. 
 For a range in $L_{\rm bol}$ from $10^{44.5}-10^{45}$~erg~s$^{-1}$ (i.e. 
assuming a minimum equal to the sum of FIR, optical, and X-rays
 luminosities, reported by \citealt{Pietsch98}, and in this work), 
we infer from the mass-luminosity-time-scale 
relation of \cite{McHardy06} that $M_{\rm BH}$ ranges from
$4\times10^{6}$--$10^{7}$~M$_{\odot}$ and the source is accreting at or 
close to the Eddington rate (assuming even higher $L_{\rm bol}$ 
implies that the AGN is a super-Eddington accretor).

 The rms-spectra obtained using light curves binned to 5~ks and
500~s show that $F_{\rm var}$
 appears to decrease above about 2~keV for both
bin timescales. Therefore the soft X-ray excess down to 0.3\,keV 
appears more variable compared to the hard energy range. This is
different from the typical rms-spectra from other Seyferts,
  which tend to show a broad peak in the variability amplitude between 0.8
and 2\,keV, and a much lower variability amplitude below 0.8 keV
(e.g., \citealt{Fabian02,Inoue03,Ponti06}). 
From the flux-flux analysis, the existing data cannot
distinguish between spectral pivoting of the continuum and a
two-component 
origin for the spectral softening, primarily because the data do not span
a broad enough flux range to see the characteristic change in gradient
expected from the power-law. However, in the case of the two-component
model being 
correct for this source, the fractional offsets measured in the
flux-flux plots increase significantly toward higher energies 
(similarly to what is observed in MCG--6-30-15),
as expected if we are witnessing a constant reflection component. 
Because there is no evidence for distant, cold material such as a
torus ($\S$4), then 
if the Compton reflection component is real, the reflection is more
likely associated 
with the accretion disc. 

\subsection{Spectral analysis: The FeK$\alpha$ line complex}

Instead of a single highly
redshifted line at about 5.4\,keV as observed in the
4\,ks May 2001 observation,  we found in the
November 2005 time-averaged observation two narrow Fe\,K$\alpha$ 
emission lines at about 6.5\,keV and 7.0\,keV. 
The former rest-frame energy is compatible within
the error bars with a line emitted by moderately ionized iron
($\sim$\ion{Fe}{xix}--\ion{Fe}{xxii}). However this line is more
  likely a blue or red peak of a relativistic line, indeed a genuine
  \ion{Fe}{xix}--\ion{Fe}{xxii} (L-ions) emission line complex would suffer
  from resonant Auger destruction \citep{Ross96,Liedahl05}.   
The latter line energy
corresponds to highly ionized iron line (\ion{Fe}{xxvi}) or to a
highly blue-shifted iron line from lower ionization matter. 
The two features can be well represented by a double-peaked line
profile from the disk with a rest-frame energy of 6.75--6.80\,keV. 
In case of an annulus {\sc diskline} profile an inner radius of about
200\,$R_{\rm g}$ is inferred.  Alternatively the \ion{Fe}{xxvi} 
line could be the signature of a strongly photoionized 
(log\,$\xi>$3), circumnuclear matter  (seen in emission)
with an iron overabundance of a factor 3, 
and/or very high column densities with $N_{\rm H,
  warm}>>10^{23}$\,cm$^{-2}$ (\citealt{Bianchi02}).  
 There is no significant evidence for line variability, however with 
the current data we also cannot reject the possibility that the lines do 
vary in response to continuum variations. 
 Further {\sl XMM-Newton} monitoring of this source can help to
establish or reject line variability at different time-scales (from
day(s) to months) and also
allow us to determine if the line profile is indeed double-peaked and
originates in the disk or if there exist independent or transient
red-/blue-shifted emission lines. 
Shorter time-scales will only be reached by higher
collecting area (and resolution at the iron K$\alpha$ energy band)
instruments such as {\sl  Constellation-X} and {\sl XEUS}. \\

In the May 2001 observation of ESO~113-G010 we observed a line feature
at 5.39~keV, which was detected at 99\% confidence (based on Monte
Carlo simulations) with EW$=233\pm140$~eV. 
We notice that a weak feature is seen at 5.4~keV near 20~ks in the
energy-time map (Fig.~9), but this line has a detection probability of
only 38\% using Monte-Carlo simulations. 
The 90\% upper limit on EW for a 5.39~keV line in the total
observation is 22~eV (Section~4.2.1).  Therefore, if the line observed
in May 2001 is real, it is transient. 
We note however that, if we treat the total data set on ESO~113-G010
as a set of independent $\sim4$~ksec segments, the chance of seeing a
99\% fluctuation in a single segment by chance becomes quite high.
However, in support of the possibility 
the 5.39~keV feature was real, we also point out that the flux and spectral
hardness in May 2001 corresponds to the extreme low and high values
observed in 2005 (see Fig.~1), which might be more favourable to
transient line formation (though we also note that such a line 
feature was not observed during the 2005 flux minimum).

In addition, as suggested by the 2001 observation,
no significant constant narrow 6.4\,keV Fe\,K${\alpha}$
line is observed with an upper-limit of $<32$\,eV, hence excluding any
dominant emission from 
distant cold matter such as a torus. For a neutral solar abundance
reflector subtending  
$2\pi$\,sr$^{-1}$ to the line of sight, illuminated by a $\Gamma=1.9$
continuum and  
inclined at 45$^{\circ}$, the expected iron line equivalent width is
$\sim125$\,eV  
with respect to the primary continuum \citep{GF91}. 
Thus in ESO\,113-G010 any cold, Compton-thick reprocessor outside of the 
line of sight is likely to subtend a much smaller solid angle, of the 
order $<\pi/2$\,sr$^{-1}$. 

 The optical classification as a Seyfert type 1.8 of ESO\,113-G010 (observed in
 November 1996; \citealt{Pietsch98}) suggests the presence of  a reddened
broad line region (BLR). For a reddening of E(B-V) = 0.8--1.0, typical for a Seyfert
 1.8  \citep{Veron06}, the hydrogen column density would be equal to or greater than about
 4--6$\times$10$^{21}$\,cm$^{-2}$ (assuming a dust-to-gas ratio equal to
 or smaller than the Galactic value; \citealt{Bohlin78}).
However, during the present 2005 {\sl XMM-Newton} observation (also suggested by the 2001
 observation), instead of an intrinsic  neutral
 absorption (in addition to the Galactic
 column density value of 2.74 $\times$ 10$^{20}$\,cm$^{-2}$) 
expected about below 1\,keV, a soft excess is observed. 
A resolution of this apparent discrepancy, without the need to invoke
 a very high dust-to-gas ratio is 
either that dust that obscures the BLR clouds could lie out of the direct
 line of sight toward the X-ray emitter; or dust responsible for the BLR
reddening could be mixed in with the warm absorber-emitter medium
 detected in the RGS  spectrum. Alternatively, the apparent disagreement
 between the (non-simultaneous) optical and X-ray classifications
 could be explained 
 by a time variability of the optical and/or X-ray absorbing properties
(e.g., \object{H1320+551}: \citealt{Barcons03}).  

\begin{acknowledgements}
The XMM-Newton project is an ESA Science Mission with instruments
and contributions directly funded by ESA Member States and the
USA (NASA). 
We thank the anonymous referee for fruitful comments and suggestions.
\end{acknowledgements}

\appendix

\section{Monte-Carlo simulations}\label{sec:mc}

\subsection{Method}
We carried out rigorous tests of the significance of the 
lines in the time-averaged spectrum and in each segment 
using Monte Carlo simulations; see \cite{P04b} 
 and \cite{Markowitz06} for full details. For the null hypothesis, we
assumed that the spectrum is simply an absorbed power-law continuum, 
with the same parameters as the absorbed power-law model
fitted to the real data in each segment. 
We used the XSPEC FAKEIT command to create 
1000 fake EPIC-pn spectra corresponding to this model, with
photon statistics appropriate for each exposure
(38.6, 24.0, and 26.9 
ks for segments 1, 2 and 3, respectively), and grouped each
spectrum to a maximum of 20 counts per bin. 
For each faked spectrum, we re-fit the null hypothesis model,
yielding a ``modified'' null hypothesis model. We ran 
FAKEIT a second time using this re-fit model, again 
with the appropriate exposure time; this process accounts
for the uncertainty in the null hypothesis model itself.
For each new spectrum, we re-fit the null hypothesis model,
recorded an initial $\chi^2$ value, and then added a Gaussian
component to the fit, as described below. The line centroid energy
was constrained to the 4.5--8.0 keV range, and the line
normalization was allowed to be positive or negative.
We stepped the Gaussian centroid energy 
over this range in increments of 0.1 keV,
fitting separately each time to ensure the lowest $\chi^{2}$ value was found. 
For each spectrum, we compared the lowest
$\chi^2$ value with that from the null
hypothesis fit, to obtain 1000 simulated
values of the $\Delta\chi^{2}$, 
which we used to construct a cumulative frequency distribution
of the $\Delta\chi^{2}$ expected for a blind line search in the
4.5--8.0 keV range, assuming the null hypothesis of a simple power-law
with no line is correct. 

\subsection{Results for the time-averaged spectrum}\label{app:average}

   The observed 6.5 and 7.0 keV lines had $\Delta\chi^2$ values of
   13.2 and 19.0, significant at 97.4$\%$ and 99.7$\%$ confidence,
   respectively, according to the simulations. However, since we
   have detected two lines that are close in energy, we can
   quantify their detection significance as a pair. The likelihood that
   both lines are due to photon noise is (1--0.974)(1--0.997) =
   8$\times$10$^{-5}$, i.e., the pair is significant at just over
   99.99$\%$ confidence. As an alternative quantification of the
   detection significance of the pair, we ran simulations
   which tested for the presence of a pair of narrow Gaussians
   with a fixed energy separation of 0.50 keV and with both line
   intensities left free. The observed $\Delta\chi^2$ value
   of 33.2 is significant at $>$99.9$\%$ confidence. However, as there
   was no {\sl a priori} expectation for the value of the energy separation,
   using such a line profile tests only a small subset of the
   full range of line pair profiles possible, and so this
   probability should be treated as an upper limit.
   Assuming the annulus profile is correct, the observed
   $\Delta\chi^2$ value of 28.8 corresponds to a detection
   probability of $>$99.9$\%$.

\subsection{Results for the time-resolved sub-spectra}\label{app:trs}

    Segment 1 (0--40ks):  The observed 6.51 keV line had a $\Delta\chi^2$ of
    9.4, significant at 85.8$\%$ according to the Monte
    Carlo simulations. The observed 7.03 keV line has a $\Delta\chi^2$ of
    4.2, significant at only 18.6$\%$ confidence, i.e., we
    cannot significantly rule out the hypothesis that the 7.03 keV
    feature is due to photon noise. However, the likelihood that
    {\sl both} lines are due to photon noise, calculated in the
    same manner as for the time-averaged spectrum, is
    0.116, i.e., the pair of lines is significant at 89.4$\%$.
    Performing Monte Carlo simulations assuming two lines with a fixed
    energy separation yields (an upper limit of) 92.0$\%$
    confidence.\\
    Segment 2 (40-70\,ks): The simulations suggest the broad Gaussian 7.0\,keV 
    line, with an observed $\Delta\chi^2$ of 17.9,
    is significant at 99.7$\%$ confidence. \\
    Segment 3 (70--100\,ks): The simulations suggest the broad
    Gaussian 6.7\,keV  
    line, with an observed $\Delta\chi^2$ of 13.9, for only 3
    additional parameters is significant at 99.9$\%$ confidence. 


\begin{thebibliography}{39}
\expandafter\ifx\csname natexlab\endcsname\relax\def\natexlab#1{#1}\fi

\bibitem[{{Anders} \& {Grevesse}(1989)}]{An89}
{Anders}, E. \& {Grevesse}, N. 1989, \gca, 53, 197

\bibitem[{{Barcons} {et~al.}(2003){Barcons}, {Carrera}, \&
  {Ceballos}}]{Barcons03}
{Barcons}, X., {Carrera}, F.~J., \& {Ceballos}, M.~T. 2003, \mnras, 339, 757

\bibitem[{{Bianchi} \& {Matt}(2002)}]{Bianchi02}
{Bianchi}, S. \& {Matt}, G. 2002, \aap, 387, 76

\bibitem[{{Bohlin} {et~al.}(1978){Bohlin}, {Savage}, \& {Drake}}]{Bohlin78}
{Bohlin}, R.~C., {Savage}, B.~D., \& {Drake}, J.~F. 1978, \apj, 224, 132

\bibitem[{{Della Ceca} {et~al.}(2005){Della Ceca}, {Ballo}, {Braito}, \&
  {Maccacaro}}]{DellaCeca05}
{Della Ceca}, R., {Ballo}, L., {Braito}, V., \& {Maccacaro}, T. 2005, \apj,
  627, 706

\bibitem[{{Dickey} \& {Lockman}(1990)}]{DL90}
{Dickey}, J.~M. \& {Lockman}, F.~J. 1990, \araa, 28, 215

\bibitem[{{Dov{\v c}iak} {et~al.}(2004){Dov{\v c}iak}, {Bianchi}, {Guainazzi},
  {Karas}, \& {Matt}}]{Dovciak04}
{Dov{\v c}iak}, M., {Bianchi}, S., {Guainazzi}, M., {Karas}, V., \& {Matt}, G.
  2004, \mnras, 350, 745

\bibitem[{{Fabian} {et~al.}(1989){Fabian}, {Rees}, {Stella}, \&
  {White}}]{Fabian89}
{Fabian}, A.~C., {Rees}, M.~J., {Stella}, L., \& {White}, N.~E. 1989, \mnras,
  238, 729

\bibitem[{{Fabian} {et~al.}(2002){Fabian}, {Vaughan}, {Nandra}, {Iwasawa},
  {Ballantyne}, {Lee}, {De Rosa}, {Turner}, \& {Young}}]{Fabian02}
{Fabian}, A.~C., {Vaughan}, S., {Nandra}, K., {et~al.} 2002, \mnras, 335, L1

\bibitem[{{George} \& {Fabian}(1991)}]{GF91}
{George}, I.~M. \& {Fabian}, A.~C. 1991, \mnras, 249, 352

\bibitem[{{Inoue} \& {Matsumoto}(2003)}]{Inoue03}
{Inoue}, H. \& {Matsumoto}, C. 2003, \pasj, 55, 625

\bibitem[{{Jim{\'e}nez-Bail{\'o}n} {et~al.}(2005){Jim{\'e}nez-Bail{\'o}n},
  {Piconcelli}, {Guainazzi}, {Schartel}, {Rodr{\'{\i}}guez-Pascual}, \&
  {Santos-Lle{\'o}}}]{Ji05}
{Jim{\'e}nez-Bail{\'o}n}, E., {Piconcelli}, E., {Guainazzi}, M., {et~al.} 2005,
  \aap, 435, 449

\bibitem[{{Liedahl}(2005)}]{Liedahl05}
{Liedahl}, D.~A. 2005, in American Institute of Physics Conference Series, Vol.
  774, X-ray Diagnostics of Astrophysical Plasmas: Theory, Experiment, and
  Observation, ed. R.~{Smith}, 99--108

\bibitem[{{Markowitz} {et~al.}(2003){Markowitz}, {Edelson}, {Vaughan},
  {Uttley}, {George}, {Griffiths}, {Kaspi}, {Lawrence}, {McHardy}, {Nandra},
  {Pounds}, {Reeves}, {Schurch}, \& {Warwick}}]{Markowitz03}
{Markowitz}, A., {Edelson}, R., {Vaughan}, S., {et~al.} 2003, \apj, 593, 96

\bibitem[{{Markowitz} {et~al.}(2006){Markowitz}, {Reeves}, \&
  {Braito}}]{Markowitz06}
{Markowitz}, A., {Reeves}, J.~N., \& {Braito}, V. 2006, \apj, 646, 783

\bibitem[{{McHardy} {et~al.}(2005){McHardy}, {Gunn}, {Uttley}, \&
  {Goad}}]{McHardy05}
{McHardy}, I.~M., {Gunn}, K.~F., {Uttley}, P., \& {Goad}, M.~R. 2005, \mnras,
  359, 1469

\bibitem[{{McHardy} {et~al.}(2006){McHardy}, {Koerding}, {Knigge}, {Uttley}, \&
  {Fender}}]{McHardy06}
{McHardy}, I.~M., {Koerding}, E., {Knigge}, C., {Uttley}, P., \& {Fender},
  R.~P. 2006, \nat, 444, 730

\bibitem[{{McHardy} {et~al.}(2004){McHardy}, {Papadakis}, {Uttley}, {Page}, \&
  {Mason}}]{McHardy04}
{McHardy}, I.~M., {Papadakis}, I.~E., {Uttley}, P., {Page}, M.~J., \& {Mason},
  K.~O. 2004, \mnras, 348, 783

\bibitem[{{Palmeri} {et~al.}(2003){Palmeri}, {Mendoza}, {Kallman}, {Bautista},
  \& {Mel{\'e}ndez}}]{Palmeri03}
{Palmeri}, P., {Mendoza}, C., {Kallman}, T.~R., {Bautista}, M.~A., \&
  {Mel{\'e}ndez}, M. 2003, \aap, 410, 359

\bibitem[{{Piconcelli} {et~al.}(2005){Piconcelli}, {Jimenez-Bail{\' o}n},
  {Guainazzi}, {Schartel}, {Rodr{\'{\i}}guez-Pascual}, \& {Santos-Lle{\'
  o}}}]{Pi05}
{Piconcelli}, E., {Jimenez-Bail{\' o}n}, E., {Guainazzi}, M., {et~al.} 2005,
  \aap, 432, 15

\bibitem[{{Pietsch} {et~al.}(1998){Pietsch}, {Bischoff}, {Boller},
  {Doebereiner}, {Kollatschny}, \& {Zimmermann}}]{Pietsch98}
{Pietsch}, W., {Bischoff}, K., {Boller}, T., {et~al.} 1998, \aap, 333, 48

\bibitem[{{Pietsch} {et~al.}(2005){Pietsch}, {Freyberg}, \&
  {Haberl}}]{Pietsch05}
{Pietsch}, W., {Freyberg}, M., \& {Haberl}, F. 2005, \aap, 434, 483

\bibitem[{{Ponti} {et~al.}(2006){Ponti}, {Miniutti}, {Cappi}, {Maraschi},
  {Fabian}, \& {Iwasawa}}]{Ponti06}
{Ponti}, G., {Miniutti}, G., {Cappi}, M., {et~al.} 2006, \mnras, 368, 903

\bibitem[{{Porquet} {et~al.}(2004{\natexlab{a}}){Porquet}, {Reeves}, {O'Brien},
  \& {Brinkmann}}]{P04a}
{Porquet}, D., {Reeves}, J.~N., {O'Brien}, P., \& {Brinkmann}, W.
  2004{\natexlab{a}}, \aap, 422, 85

\bibitem[{{Porquet} {et~al.}(2004{\natexlab{b}}){Porquet}, {Reeves}, {Uttley},
  \& {Turner}}]{P04b}
{Porquet}, D., {Reeves}, J.~N., {Uttley}, P., \& {Turner}, T.~J.
  2004{\natexlab{b}}, \aap, 427, 101

\bibitem[{{Ross} {et~al.}(1996){Ross}, {Fabian}, \& {Brandt}}]{Ross96}
{Ross}, R.~R., {Fabian}, A.~C., \& {Brandt}, W.~N. 1996, \mnras, 278, 1082

\bibitem[{{Str{\" u}der} {et~al.}(2001){Str{\" u}der}, {Briel}, {Dennerl},
  {Hartmann}, {Kendziorra}, {Meidinger}, {Pfeffermann}, {Reppin}, {Aschenbach},
  {Bornemann}, {Br{\" a}uninger}, {Burkert}, {Elender}, {Freyberg}, {Haberl},
  {Hartner}, {Heuschmann}, {Hippmann}, {Kastelic}, {Kemmer}, {Kettenring},
  {Kink}, {Krause}, {M{\" u}ller}, {Oppitz}, {Pietsch}, {Popp}, {Predehl},
  {Read}, {Stephan}, {St{\" o}tter}, {Tr{\" u}mper}, {Holl}, {Kemmer},
  {Soltau}, {St{\" o}tter}, {Weber}, {Weichert}, {von Zanthier},
  {Carathanassis}, {Lutz}, {Richter}, {Solc}, {B{\" o}ttcher}, {Kuster},
  {Staubert}, {Abbey}, {Holland}, {Turner}, {Balasini}, {Bignami}, {La
  Palombara}, {Villa}, {Buttler}, {Gianini}, {Lain{\' e}}, {Lumb}, \&
  {Dhez}}]{Str01}
{Str{\" u}der}, L., {Briel}, U., {Dennerl}, K., {et~al.} 2001, \aap, 365, L18

\bibitem[{{Taylor} {et~al.}(2003){Taylor}, {Uttley}, \& {McHardy}}]{Taylor03}
{Taylor}, R.~D., {Uttley}, P., \& {McHardy}, I.~M. 2003, \mnras, 342, L31

\bibitem[{{Turner} {et~al.}(2001){Turner}, {Abbey}, {Arnaud}, {Balasini},
  {Barbera}, {Belsole}, {Bennie}, {Bernard}, {Bignami}, {Boer}, {Briel},
  {Butler}, {Cara}, {Chabaud}, {Cole}, {Collura}, {Conte}, {Cros}, {Denby},
  {Dhez}, {Di Coco}, {Dowson}, {Ferrando}, {Ghizzardi}, {Gianotti}, {Goodall},
  {Gretton}, {Griffiths}, {Hainaut}, {Hochedez}, {Holland}, {Jourdain},
  {Kendziorra}, {Lagostina}, {Laine}, {La Palombara}, {Lortholary}, {Lumb},
  {Marty}, {Molendi}, {Pigot}, {Poindron}, {Pounds}, {Reeves}, {Reppin},
  {Rothenflug}, {Salvetat}, {Sauvageot}, {Schmitt}, {Sembay}, {Short},
  {Spragg}, {Stephen}, {Str{\" u}der}, {Tiengo}, {Trifoglio}, {Tr{\" u}mper},
  {Vercellone}, {Vigroux}, {Villa}, {Ward}, {Whitehead}, \& {Zonca}}]{Turner01}
{Turner}, M.~J.~L., {Abbey}, A., {Arnaud}, M., {et~al.} 2001, \aap, 365, L27

\bibitem[{{Turner} {et~al.}(2004){Turner}, {Kraemer}, \& {Reeves}}]{JTurner04}
{Turner}, T.~J., {Kraemer}, S.~B., \& {Reeves}, J.~N. 2004, \apj, 603, 62

\bibitem[{{Turner} {et~al.}(2006){Turner}, {Miller}, {George}, \&
  {Reeves}}]{Tu06}
{Turner}, T.~J., {Miller}, L., {George}, I.~M., \& {Reeves}, J.~N. 2006, \aap,
  445, 59

\bibitem[{{Turner} {et~al.}(2002){Turner}, {Mushotzky}, {Yaqoob}, {George},
  {Snowden}, {Netzer}, {Kraemer}, {Nandra}, \& {Chelouche}}]{Tu02}
{Turner}, T.~J., {Mushotzky}, R.~F., {Yaqoob}, T., {et~al.} 2002, \apjl, 574,
  L123

\bibitem[{{Uttley} \& {McHardy}(2005)}]{Uttley05}
{Uttley}, P. \& {McHardy}, I.~M. 2005, \mnras, 363, 586

\bibitem[{{Vaughan} {et~al.}(2003{\natexlab{a}}){Vaughan}, {Edelson},
  {Warwick}, \& {Uttley}}]{Vaughan03b}
{Vaughan}, S., {Edelson}, R., {Warwick}, R.~S., \& {Uttley}, P.
  2003{\natexlab{a}}, \mnras, 345, 1271

\bibitem[{{Vaughan} \& {Fabian}(2004)}]{Vaughan04}
{Vaughan}, S. \& {Fabian}, A.~C. 2004, \mnras, 348, 1415

\bibitem[{{Vaughan} {et~al.}(2003{\natexlab{b}}){Vaughan}, {Fabian}, \&
  {Nandra}}]{Vaughan03a}
{Vaughan}, S., {Fabian}, A.~C., \& {Nandra}, K. 2003{\natexlab{b}}, \mnras,
  339, 1237

\bibitem[{{V{\'e}ron-Cetty} \& {V{\'e}ron}(2006)}]{Veron06}
{V{\'e}ron-Cetty}, M.-P. \& {V{\'e}ron}, P. 2006, \aap, 455, 773

\bibitem[{{Wilms} {et~al.}(2000){Wilms}, {Allen}, \& {McCray}}]{Wilms00}
{Wilms}, J., {Allen}, A., \& {McCray}, R. 2000, \apj, 542, 914

\bibitem[{{Yaqoob} {et~al.}(2003){Yaqoob}, {George}, {Kallman}, {Padmanabhan},
  {Weaver}, \& {Turner}}]{Yaqoob03}
{Yaqoob}, T., {George}, I.~M., {Kallman}, T.~R., {et~al.} 2003, \apj, 596, 85

\end{thebibliography}
\end{document}